# Ultra-High Brightness Electron Beams from Very-High Field Cryogenic Radiofrequency Photocathode Sources

J.B. Rosenzweig[1], A. Cahill[1], B. Carlsten[4], G. Castorina[2], M. Croia[2], C. Emma[3], A. Fukusawa[1], B. Spataro[2], D. Alesini[2], V. Dolgashev[3], M. Ferrario[2], G. Lawler[1], R. Li[3], C. Limborg[3], J. Maxson[1], P. Musumeci[1], R. Pompili[2], S. Tantawi[3], O. Williams[1]

[1]*Dept. of Physics and Astronomy, University of California, Los Angeles, California, USA*
[2]*Laboratori Nazionali di Frascati, INFN, Via E. Fermi, Frascati, Italia*
[3]*SLAC National Accelerator Laboratory, Menlo Park, California, USA*
[4]*Los Alamos National Laboratory, Los Alamos, New Mexico, USA*

**Abstract**
Recent investigations of RF copper structures operated at cryogenic temperatures performed by a SLAC-UCLA collaboration have shown a dramatic increase in the maximum surface electric field, to 500 MV/m. We examine use of these fields to enable very high field cryogenic photoinjectors that can attain over an order of magnitude increase in peak electron beam brightness. We present beam dynamics studies relevant to X-ray FEL injectors, using start-to-end simulations that show the high brightness and low emittance of this source enables operation of a compact FEL reaching a photon energy of 80 keV. The preservation of beam brightness in compression, exploiting micro-bunching techniques is discussed. While the gain in brightness at high field is due to increase of the emission current density, further increases in brightness due to lowering of the intrinsic cathode emittance in cryogenic operation are also enabled. While the original proposal for this type of cryogenic, ultra-high field photoinjector has emphasized S-band designs, there are numerous potential advantages that may be conferred by operation in C-band. We examine issues related to experimental implementation in C-band, and expected performance of this type of device in a future hard X-ray FEL such as MaRIE.

## Introduction

The introduction of fundamentally higher brightness electron sources, facilitated by the introduction of the pulsed high-field radiofrequency (RF) photoinjector over 25 years ago [1,2], has changed the trajectory of science. These sources have permitted the production of intense, cold, relativistic electron beams with ultra-fast time structures, in the earliest days at the picosecond scale, but now reaching the femtosecond level. Such RF photoinjector sources, based on short-pulse laser excitation of a photocathode embedded in a high-field RF accelerator environment, have proven to be essential experimental instruments in beam physics, enabling many high impact applications, in particular allowing a wide variety of powerful new ultra-fast light sources. This new generation of photon-based imaging instruments prominently includes X-ray free-electron lasers (XFELs) [3,4,5,6,7], which are now diffused worldwide. The X-ray FEL has, since its introduction, revolutionized X-ray-based imaging via the introduction of coherence in light with wavelengths extending to the Angstrom level. With fsec pulses, this instrument yields detailed information about the behavior and structure of atomic-molecular systems at their characteristic spatial and temporal scales, permitting ultra-fast four-dimensional imaging.

In this list of applications of high brightness electron beams, one stands out in terms of impact — the central role played in creating the lasing medium in the XFEL, which has introduced a new, 4th-generation of X-ray light sources [8,9,10], enabling the properties of coherence and ultrafast time scales into high-flux hard X-ray sources. The achieving of self-amplified spontaneous emission (SASE) FEL operation using a vigorous high gain regime based on high brightness electron beams in turn has yielded X-ray light sources with approximately ten orders of magnitude increase in photon spectral brightness. These extremely bright, coherent light sources have introduced high impact methods in X-ray based science [11]. Beyond the LCLS, SACLA, PAL and the European XFEL, the hard X-ray frontier is expected to be pushed in wavelength, to the well sub-Å level, in the nascent MaRIE project at the Los Alamos National Laboratory.

High electron beam brightness is essential to the physics underpinning the X-ray FEL; in the absence of an increase of over an order of magnitude in electron beam brightness over the then state-of-the-art, the LCLS would not have been possible. This beam brightness, being the ratio of the peak current to the 4D transverse phase space area, $B_e = 2I/\varepsilon_x^2$, enters into the physics of the FEL in a variety of ways, some direct and others through emphasis on the individual factors $I$ and $\varepsilon_x$. These include control of the FEL gain length through the Pierce parameter [12], $L_g \propto \rho^{-1} \propto B_e^{-1/3}$. Further, the efficiency of energy extraction from the electron beam in FEL saturation is proportional to $\eta \equiv U_{FEL}/U_{e-} \cong \rho$; tapering schemes to enhance the efficiency are more effective when the nominal saturation power is high. The peak FEL power is of proportional to $I$, emphasizing that charge and/or current cannot be sacrificed in pursuing large photon flux experiments such as single molecule imaging. On the other hand, at reduced charge, the emittance and pulse length may be minimized, allowing access to new FEL regimes, such a sub-fs and/or single spike operation. Reductions of $\varepsilon_x$ also bring advantages, as with significantly small emittances, and access to shorter wavelength FEL radiation, an option of high interest in dense, high-Z matter imaging such as at MaRIE.

**Source Brightness and RF Field at Emission**

With the central role played by electron beam brightness and emittance in determining the performance of the XFEL and other applications, increasing $B_e$ and lowering $\epsilon_n$ has taken on increased urgency. This brightness cannot be increased beyond the intrinsic value obtained at the cathode. Compactly, the intrinsic beam brightness is inversely proportional the effective cathode temperature $T_c$ [13]. Here the parameters $k_B$ and $m_e c^2$ indicate the Boltzmann constant and the electron rest energy, respectively. In the brightness definition the current $I$ is divided by the transverse normalized emittance squared $\varepsilon_n^2$; we recast this in terms of emission current density $J_z$ and the temperature $T_c$, as both of these parameters can be possibly improved.

For the sub-ps emission from metallic surfaces $k_B T_c$ is proportional with a factor near unity to the difference between the laser photon energy $h\nu$ and the metal's work function $W$; we further assumed that the transverse and longitudinal temperatures are similar at emission. We note that this assertion concerns scenarios where the photocathode ambient *material* (internal electron) temperature is ignorable. Thus one may improve $B_0$ (or $B_{0,6D}$) by either increasing $J_z$ or lowering $T_c$.

Here we concentrate on the strong – orders of magnitude – increase in the beam current density permitted by very high field photoinjectors. It also implies that one must transport and compress the beam without emittance dilution. The advantages of high field operation are explicitly noted from the expression one may write on the maximum current density obtained from a photocathode in the 1D space-charge (longitudinal blowout, regime, per the discussions in Ref. 14 ) limit is $J_z = \dfrac{ec\varepsilon_0}{m_e c^2}\left(E_0 \sin\varphi_0\right)^2$ where $E_0 \sin\varphi_0$ is the launch field at the photocathode. We can employ this expression to estimate the associated 1D intrinsic limit on beam brightness, $B_0 = J_z = \dfrac{2ec\varepsilon_0}{k_B T_c}\left(E_0 \sin\varphi_0\right)^2$. This scaling is modified somewhat when one considers 3D effects on space charge at emission. If one uses a beam much narrower than its length, the scaling of current density with field is proportional to $\left(E_0 \sin\varphi_0\right)^{3/2}$, giving excellent performance, but only for relatively low charge beams. The conclusion is there are potentially very large advantages conferred to operation of RF photoinjectors at large launch field.

**Achieving and Applying High Fields in Standing Wave Cavities**

In this regard, a new paradigm for photoinjector design and realization has been proposed recently, that takes as its point of departure the breakthrough work in the development of cryogenically-cooled Cu RF structures carried out at SLAC in recent years. In tests on X-band structures, enhanced $Q$ and significantly improved gradients at 45 K, corresponding to nearly 500 MV/m surface fields before breakdown, have been demonstrated [15]. The advantage in $Q$ arises from the reduction surface dissipation associated with the anomalous skin effect (ASE) in Cu at low temperatures. This lower dissipation implies that there is diminished thermal stress on the surface. At low temperatures, the material yield strength increases dramatically. In combination with lowered RF dissipation, the peak surface fields achieved in X-band tests without breakdown reach 500 MV/m. This is a qualitative change in behavior, in the sense that there seems to be a threshold for breakdown below which the statistical probability of observing breakdown is vanishingly small. Indeed, the limit on utility of these gradients observed in these tests is presently found in dark current-loading of the cavities above ~300 MV/m peak surface field [16].

Based on this leap forward in high field performance, the UCLA-SLAC-LNF collaboration contributing also to this article, have analyzed a scenario that applies cryogenic operation of Cu cavities, as applied to an advanced RF photocathode gun in S-band [17]. This discussion concentrates on development of a 1.45 cell gun operated at ~27 K, with a $Q$ value enhanced by a factor of up to 5 and a peak electric field on the cathode of 250 MV/m. Detailed simulations indicate that one may achieve $\varepsilon_n$ =40 nm at 200 pC, in a 10 ps beam pulse. This solution thus has advantages in beam dynamics, as the peak field at photo-emission is unprecedentedly high, at ~240 MV/m.

The key drawback of the S-band approach concerns RF pulse length; even using a highly over-coupled system and 50 MW of input power, the RF pulse may not be chosen below ~0.9 μsec. To avoid such a long pulse length one may operate at high frequency, as the scaling of fill time as ~$\omega^{-3/2}$ indicates. Further, considering a constant $E_o$, the power needed to drive a structure of scaled geometry is also smaller by ~$\omega^{-3/2}$. Thus higher $\omega$ mitigates both dissipation and power usage/cooling load. In the S-band case studied, a 120 Hz, photoinjector dissipates ~500 W at cryo-temperatures, necessitating a cryo-cooling system using 25 kW. A C- or X-band system would be much less demanding. Further, at these $E_o$ values, integrated dark current is exacerbated by long RF pulses [18]. Here, we are interested in exploring C-band systems, that can operate with pulses ~300 nsec.

**A Scaled C-band Photoinjector**

In experimental testing at high frequency, guns up to 17 GHz have been tested [19]. Few X-Band RF guns have been tested at high power thus far [20], and predicted beam characteristics have been recently demonstrated on the two first X-Band guns [21,22]. At charges relevant to FEL operation, bunch lengths as short as 400 fs rms have been achieved for 100 pC bunches, with $\varepsilon_n = 0.7$ mm-mrad. These parameters do not approach those promised by the S-band case in Ref. [**Error! Bookmark not defined.**], with obtained brightness lower by many orders of magnitude. In the most recent studies [22], a damaged cathode prevented experimental demonstration of the predicted $\varepsilon_n = 0.2$ mm-mrad (95%) at 200 MV/m. The potential of an X-Band gun remains to be demonstrated. The main problem in X-band photoinjectors is that they operate at a low value of the normalized vector potential $\alpha_{RF} = eE_0 / \omega_{RF} m_e c$ [23], equivalent to an S-band case with <50 MV/m peak field. This leads to beam dynamics challenges; in particular, the launch phase is difficult to place near crest. This can be partially addressed by use of a shorter initial cell (0.4λ/2 instead of 0.6 λ/2 in the LCLS gun), but at the cost of degraded emittance compensation due to lack of RF focusing.

If one considers operation in C-band, however, one can directly scale state-of-the-art S-band designs to fields twice as large, as the field should scale accordingly, $E_0 \propto \omega$. In fact, a re-optimization of the current LCLS photoinjector [24] at $E_0$=120 MV/m has predicted that using 10 ps bunch lengths and 200 pC charge, an emittance of $\varepsilon_n$ = 110 nm-rad. This new scenario implies working near the cigar-beam limit, and to move the position of the post-acceleration linac from $z$=1.5 m to $z$=2.2 m.

We can immediately profit from this design work by using the well-known scaling methods developed in Ref. [25] to establish a design point at 240 MV/m, nearly identical to the value assumed for the S-band ($f_{RF}$ =2.856 GHz) cryogenic gun study of Ref. [**Error! Bookmark not defined.**] by changing the operating RF frequency to C-band ($f_{RF}$ =5.712 GHz). In this case we must also scale the focusing fields up by 2, and shrink all beamline dimensions similarly. To preserve the beam collective behavior, we must also scale all the beam dimensions $\sigma_i \propto \lambda_{RF}$ and $Q \propto \lambda_{RF}$. As a result of these well-established theoretical scaling laws the beam envelope and emittance evolution are preserved, and the emittance also is known to scale as $\varepsilon_n \propto \lambda_{RF}$, if the fields and beam dimensions are indeed scaled correctly.

This approach is validated by the GPT simulations shown in Figure 1. With 100 pC in a scaled C-band 1.6 RF gun having the same interior shape as the standard S-band device [24], and using post-acceleration (with C-band linacs operated at $E_{acc}$ = 35 MV/m) that commences at $z$=1.1 m downstream of the photocathode [26], we achieve $\varepsilon_n$ = 55 nm-rad with 20 A peak current. Thus, one obtains a factor of four increased brightness with the C-band scaled option over the re-optimized LCLS case. This example is every important, in that it shows a path to profiting from the promise of the scaling discussed in Ref. [25]. This type beam can be utilized to drive an FEL with new capabilities, as studied in Ref. [17]. In the present case, the 100 pC beam with $\varepsilon_n$ =55 mm-mrad was compressed into micro-bunches using the ESASE process to >9 kA peak current at 14 GeV. Injecting this beam into a cryogenic undulator [27] having period $\lambda_u$ =9 mm, and strength $K$=1.8, saturation is predicted within ~20 m at $\lambda$=0.155 Å (80 keV) [28], as is verified in simulation. The total energy radiated at $z$~20 m is 225 µJ in this case, as is shown in Figure 2. With this promising scenario in hand, we have begun to examine optimization of the C-band photoinjector, as appllid to a MaRIE FEL-like design. This study is motivated not only by the extension of wavelength reach by an order of magnitude with low emittance, but by having an RF structure that can be operated in very short pulses, giving the flexible format needed for MaRIE.

**C-band RF Gun Design Studies**

With the preferred maximum gradient identified as near 250 MV/m, we have proceeded to address practical designs. It was found that the scaled beam dynamics are excellent for the choice of a 1.6 cell gun, and further investigations have revealed little advantage in shortening the initial cell –the benefits of field at emission for shorter cells are counterbalanced by the improved transverse beam dynamics in this charge range. Therefore, we have kept this aspect of the current generation of S-band guns in the design, with slight modifications in the iris shape to minimize RF heating. While the beam dynamics are thus scaled nearly identically in going from S-band to C-band, we must address some issues associated with very high field operation, including coupling and solenoid design that arise due to the more compact sizes associated with the gun and focusing solenoid. We next briefly discuss these design challenges.

In the case of the gun, we have a challenge in coupling the power to the RF cavity. This coupling is chosen to be well beyond critical, $\beta$=9. This corresponds to a $\beta$~2 at room temperature, and thus the structure has an external coupling similar to current guns. We have studied a relatively standard approach to the coupling, using a mode-launcher geometry that has been symmetrized using four ports to eliminate the potentially deleterious effects of time-dependent quadrupole fields. The resultant geometry is shown in the HFSS simulation results (with resonant fields in gun displayed) of Figure 3. The gun requires just under 20 MW to achieve 250 MV/m peak field, with a fill time of 315 ns. The peak power absorbed in the cavity is 7.1

MW under these coupling conditions. The dissipated power for a mean repetition rate of 120 Hz <200 W, less than half of the S-band case. There is room in this approach to either achieve higher average repetition rate, or to mitigate the cryo-cooling requirements.

The mode launcher presents challenges for a comprehensive photoinjector, however, in that it occludes space that may needed for the post-gun focusing solenoid. In particular the width of the cylindrical waveguide is roughly that of the gun, which keeps the solenoid coils and iron yoke from approaching the axis. Given the difficulty, discussed below, of achieving the needed magnetic field, this is a notable disadvantage. One may, to avoid using this post-gun space, couple directly to the RF structure on the side, as is traditionally done in S-band. In this case, one must confront problems in managing the pulsed heating at the coupling holes or slots. With the RF dissipation lowered compared to room temperature devices by a factor of over 4, the heat deposited is similar to the LCLS at 120 MV/m. However, the structure is smaller in size, and its ability to absorb power with heating the bulk structure excessively may be questioned.

To mitigate this issue we are presently examining the coupling of the external power into the 0.6-cell operated in the $TM_{02}$ mode, using a coupler similar in design approach to that of the LCLS and other recent S-band guns. This scheme is illustrated in Figure 4, which shows the electric and magnetic fields of the mode. With coupling occurring where there is more material thickness, the heating induced locally on the coupling slot curved surfaces is no larger than that on the irises, and thus the coupler is not the most sensitive thermal region in the structure, as it is in previous S-band guns. This approach may give an advantage in solenoid design.

**C-band Gun Solenoid Design Approach**

As the beam dynamics in the case shown in Figure 1, we may scale the solenoid geometry as well. Again, we run into thermal management issues however. To appreciate this, we show in Figure 5 a simulation of the scaled performance of the solenoid for the C-band gun. To reach the required peak field of 6.2 kG, the required current density is 1200 A/cm$^2$. As this is in excess of the nominal limit for water-cooled coils in present use, we are examining cryogenic operation of the coils. At 77 deg K, assuming a RRR=200, the power dissipated in the coils is diminished from 1.2 kW at room temperature to 150 W. At 40 deg K, this is lowered further, to 14 W. Given the need for a relatively sophisticated cryostat and cryo-cooler in this system, this solution seems tenable.

**Conclusions and Future work**

This work shows the initial steps taken to the potential development of a C-band cryogenic RF photoinjector, with the application to a very hard X-ray FEL, similar in characteristics to MaRIE. With source emittances a factor of ~4 lower than those presently achieved at the LCLS at similar currents. These very bright beams are thus ideal for driving FELs with very short wavelengths. Here we have shown that one may achieve in C-band what has also been found in S-band, that an 80 keV high gain FEL is possible using this injector, final microbunching compression, and an advanced short period undulator.

Initial work on the technical realization of such an injector, and its incorporation in more mature designs, is commencing. We are still establishing the contours of cryogenic cavity performance in S-band and in C-band, and will be performing high power tests analogous to those already carried out in X-band in the coming year. We will also begin addressing the problem of dark current and its mitigation, examining high field emission of structures with coatings such as graphene. Finally, we have shown several promising approaches to the RF coupler, and are evaluating their merits, particularly in regards to interferences with solenoid optimization. These issues must be examined in the context of a first attempt at integrating the design into a a functional cryostat.

**Acknowledgments**

This work was supported by the supported by the U.S. DOE contract DE-SC0009914, the U.S. NSF Award PHY-1549132, the Center for Bright Beams and DOE/SU Contract DE-AC02-76-SF00515.

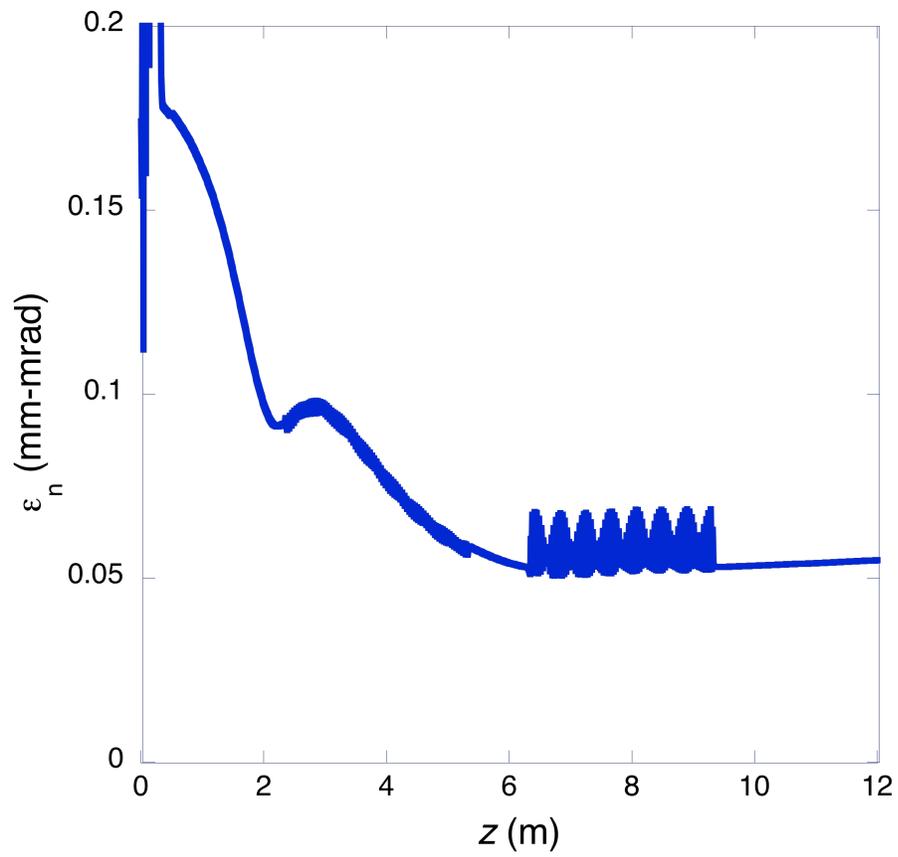

**Figure 1. Emittance evolution for C-band 1.6 cell gun RF photoinjector, with C-band post-acceleration, yielding 55 nm-rad normalized emittance.**

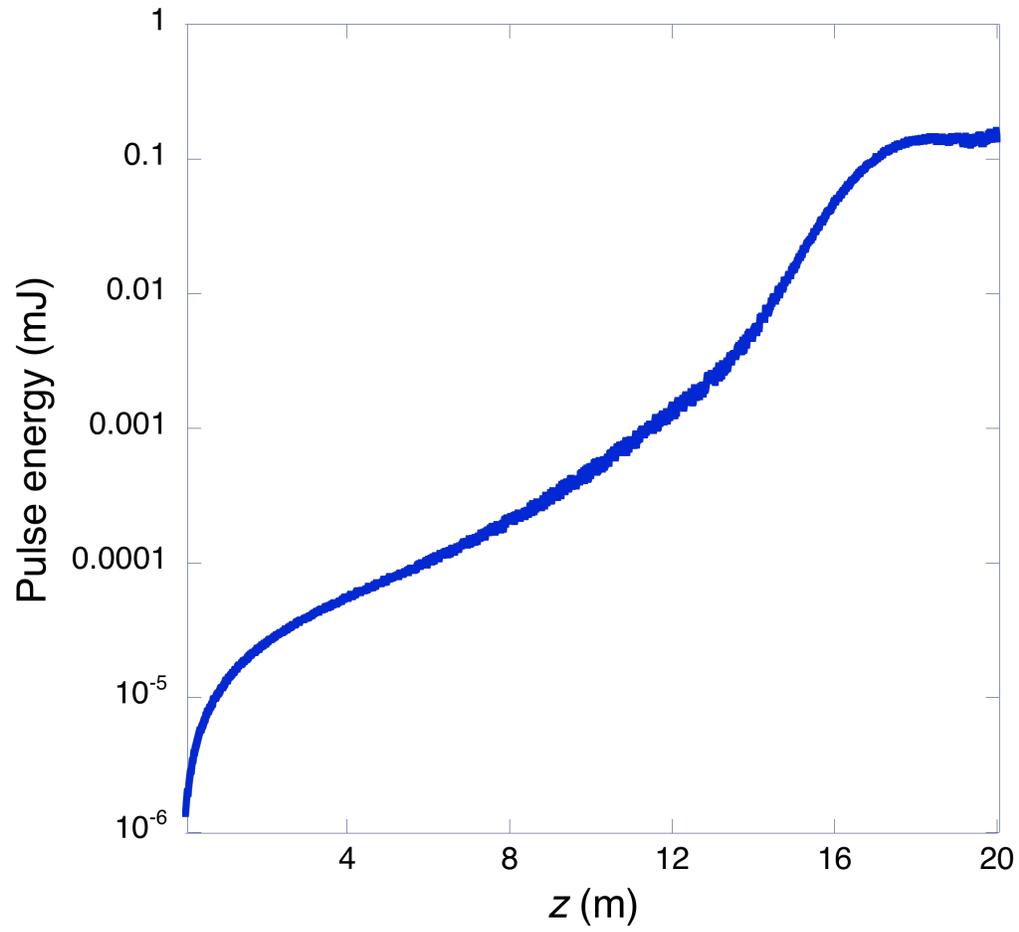

**Figure 2.** GENESIS 1.3 simulation of energy evolution of FEL performance 100 pC beam from C-band cryogenic photoinjector microbunched to 9 kA at 14 GeV and injected into 9 mm period high field undulator.

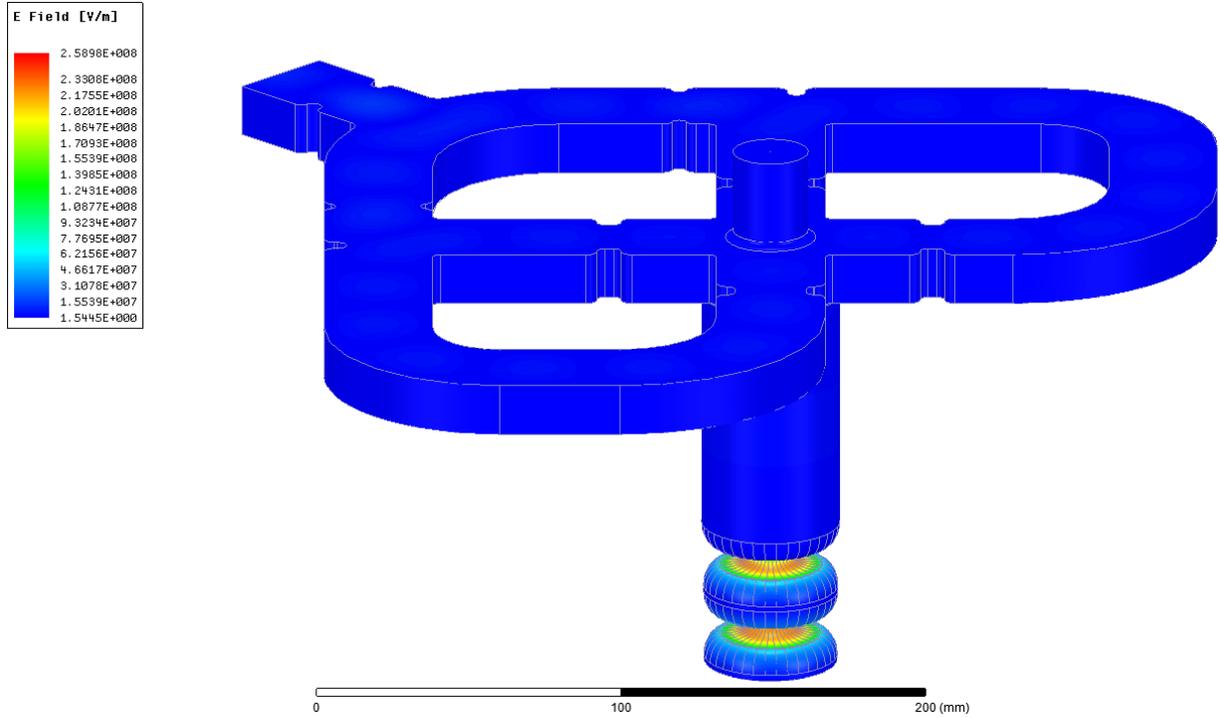

**Figure 3. External RF coupling network and 1.6 cell C-band gun, from HFSS simulation, with color map showing electric field a surface (peak 250 MV/m)**

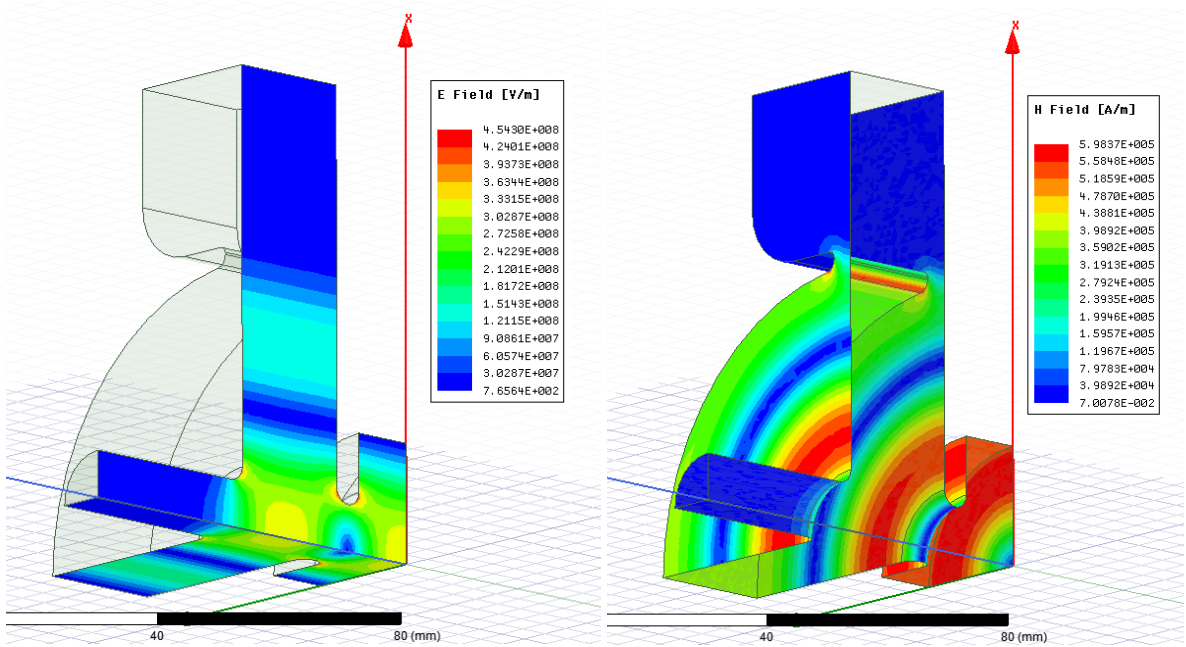

**Figure 4. Example C-band gun cathode cell operated in $TM_{02}$ mode, with external coupling, from HFSS simulation. On the left, electric fields are shown on symmetry plane cuts of the cavity; on the right, the magnetic fields are shown on the copper surfaces of the cavity.**

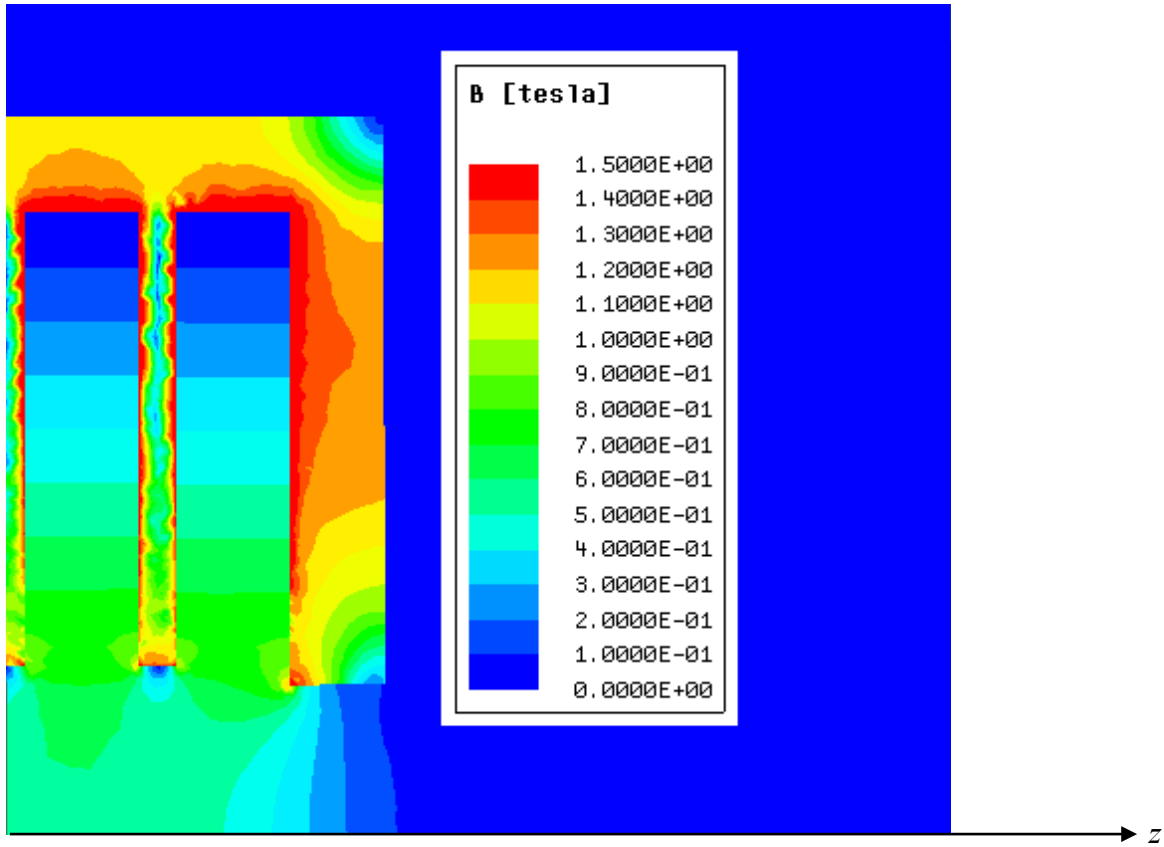

**Figure 5.** Maxwell simulation of solenoid for C-band gun, showing one-half-length of axis of rotation about beam axis ($z$). The current density is 1200 A/cm$^2$, and the half-length of the solenoid iron is 13 cm.

# References

1. J.S. Fraser, R.L. Sheffield, E.R. Gray, G.W. Rodenz, *IEEE Trans. on Nuclear Science*, 32, 1791 (1985)
2. B.E. Carlsten, *Nucl. Instr. and Meth. A*, 285 313 (1989)
3. C. Pellegrini. *Eur. Phys. J. H* **37,** 659-708 (2012).
4. P. Emma, *et al.*, *Nature Photonics* 4, 641 (2010)
5. Phillip Ball, *Nature* **548,** 507 (2017)
6. Zhirong Huang, Ingolf Lindau, *Nature Photonics* 6, 505–506 (2012)
7. Heung-Sik Kang, *et al*., *Nature Photonics* **11**, 708 (2017)
8. J. Andruszkow, et al., *Phys. Rev. Lett*. 85, 3825 (2000
9. D. Pile. *Nature Photonics* **8**, 82 (2014)
10. C. Pellegrini, *Physica Scripta*, **2016**, T169 (2017)
11. Henry N. Chapman, *et al., Nature* 470, 73 (2011)
12. R. Bonifacio, C. Narducci and C. Pellegrini, *Opt. Commun.* 50, 373 (1984).
13. B. J. Claessens*, et al., Phys. Rev. Lett*. **95,** 164801 (2005)
14. O. J. Luiten *et al.*, *Phys. Rev. Lett*. **93**, 094802 (2004).
15. A. D. Cahill *et al.* "Ultra High Gradient Breakdown Rates in X-Band Cryogenic Normal Conducting RF Accelerating Cavities", Proceedings of International Part. Accel. Conf. 2017 (JaCOW, 2017).
16. "Dynamically Changing Quality Factor in a Cryogenic Copper Cavity", A. D. Cahill and J. B. Rosenzweig, V. A. Dolgashev, Z. Li, S. G. Tantawi and S. Weathersby, submitted to *Physical Review Letters*.
17. J. Rosenzweig, et al, "Next Generation High Brightness Electron Beams From Ultra-High Field Cryogenic Radiofrequency Photocathode Source", submitted to *PRAB* (2016)*,* http://arxiv.org/abs/1603.01657
18. E.E. Wisniewski "Cs2Te Photocathode performance in the AWA High Charge High Gradient Drive Gun", IPAC 2015 , http://accelconf.web.cern.ch/AccelConf/IPAC2015/papers/wepty013.pdf
19. W. J. Brown, et al., Nucl. Instrum. Methods A 425, 459 (1999).
20. Y. Taira, et al. Nucl. Instrum. Methods A 729, 605 (2013)
21. R. Marsh, et al., PRST-AB 15, 102001 (2012)
22. C. Limborg et al. PR-AB, Vol 19, 053401 (2016)
23. K.J. Kim, *Nucl. Instruments and Methods A* **275,** 201 (1989)
24. R. Akre, *et al*., *Phys.Rev. ST Accel. Beams* 11 030703 (2008).
25. J.B. Rosenzweig, E. Colby, AIP Conference Proceedings, vol. 335, 1995, p. 724
26. T. Sakurai, et al., *Phys. Rev. Accel. Beams* 20, 042003 (2017)
27. F. H. O'Shea, *et al. J. Phys. B: At. Mol. Opt. Phys.* 47 234006 (2014)
28. G. Marcus, E. Hemsing, J. Rosenzweig, *Phys. Rev. ST Accel. Beams* **14**, 080702 (2011)